\documentclass[journal]{IEEEtran}

\usepackage{cite}

\ifCLASSINFOpdf
  \usepackage[pdftex]{graphicx}
  \graphicspath{{}}
\else
  \usepackage[dvips]{graphicx}
  \graphicspath{{}}
\fi

\usepackage{amsmath, amssymb}
\usepackage{array}

\ifCLASSOPTIONcompsoc
  \usepackage[caption=false,font=normalsize,labelfont=sf,textfont=sf]{subfig}
\else
  \usepackage[caption=false,font=footnotesize]{subfig}
\fi

\usepackage{fixltx2e}

\usepackage{url}

\begin{document}

\title{Experimental Validation of Ultrasound Beamforming with End-to-End Deep Learning for Single Plane Wave Imaging}

\author{Ryan~A.L.~Schoop, Gijs Hendriks, Tristan van Leeuwen, Chris L. de Korte, Felix Lucka% <-this % stops a space
\thanks{R.A.L. Schoop is with Image Guided Surgery, Department of Surgery, Netherlands Cancer Institute, Amsterdam, The Netherlands; and the Faculty of Science and Technology, University of Twente, Enschede, The Netherlands; e-mail: r.schoop@nki.nl}%
\thanks{G. A.G.M. Hendriks was with MUSIC 766 Radboudumc, Nijmegen the Netherlands}%
\thanks{T van Leeuwen is with the Computational Imaging group, Centrum Wiskunde \& Informatica, Amsterdam, The Netherlands and the Mathematical Institute, Utrecht University, Utrecht, The Netherlands}% <-this % stops a 
\thanks{C.L. de Korte is with MUSIC 766 Radboudumc, Nijmegen the Netherlands and Physics of Fluids, M3I, University of Twente, Enschede, the Netherlands}%
\thanks{F. Lucka is with the Computational Imaging group, Centrum Wiskunde $\&$ Informatica, Amsterdam, The Netherlands,  e-mail: Felix.Lucka@cwi.nl}%
}

\maketitle

\begin{abstract}
Ultrafast ultrasound imaging insonifies a medium with one or a combination of a few plane waves at different beam-steered angles instead of many focused waves. It can achieve much higher frame rates, but often at the cost of reduced image quality. 
Deep learning approaches have been proposed to mitigate this disadvantage, in particular for single plane wave imaging. Predominantly, image-to-image post-processing networks or fully learned data-to-image neural networks are used. Both construct their mapping purely data-driven and require expressive networks and large amounts of training data to perform well. 
In contrast, we consider data-to-image networks which incorporate a conventional image formation techniques as differentiable layers in the network architecture. This allows for end-to-end training with small amounts of training data. 
In this work, using $f$-$k$ migration as an image formation layer is evaluated in-depth with experimental data.
We acquired a data collection designed for benchmarking data-driven plane wave imaging approaches using a realistic breast mimicking phantom and an ultrasound calibration phantom. The evaluation considers global and local image similarity measures and contrast, resolution and lesion detectability analysis.    
The results show that the proposed network architecture is capable of improving the image quality of single plane wave images on all evaluation metrics. Furthermore, these image quality improvements can be achieved with surprisingly little amounts of training data.
\end{abstract}

\begin{IEEEkeywords}
Ultrasound, beamforming, plane wave imaging, deep learning, image enhancement.
\end{IEEEkeywords}

\IEEEpeerreviewmaketitle

\section{Introduction}

\IEEEPARstart{D}{iagnostic} imaging plays an important role in healthcare, because it allows for timely diagnosis, disease staging, treatment choice, planning, guidance and treatment follow-up \cite{SzaboThomasL2014Duii}.
Ultrasound imaging serves as a valuable diagnostic tool because of its low cost, portability, accessibility, real-time imaging capabilities, and non-invasive nature.
Focused wave insonification methods have been the standard for ultrasound imaging for many years \cite{LuijtenBen2020AUBU}.
In these focused methods the acquisition is done line-by-line, in which for each acquisition line the sub-region of the image corresponding to the acquired line is processed and reconstructed.
Consequently, the total time to form an image increases linearly with the number of acquisition lines necessary.
In practice, this limits the frame rate to roughly 20 to 50 frames per second.

Technological advances have allowed for higher frame rates to be achieved by using unfocused sonification techniques, in particular using plane wave ultrasound imaging \cite{LuijtenBen2020AUBU, AlbertiGiovanniS.2017MAoU}.
With this technique the acquisition is no longer line-by-line, but rather a single acquisition is used spanning the whole image region which is then processed and reconstructed.
A frame rate in the order of several thousand frames per second can be obtained with single plane wave imaging, which allows for highly dynamic imaging applications such as: blood flow imaging, shear wave elastography, or super resolution ultrasound localization microscopy \cite{TanterMickael2014Uiib, AlbertiGiovanniS.2017MAoU, BercoffJ2004Ssia, ErricoClaudia2015Uulm}.
Single plane wave imaging however comes with a drawback of degraded image quality in terms of resolution, contrast, and penetration depth when compared to conventional focused ultrasound imaging.
Image quality can be restored and even improved by coherently compounding the ultrasound signals  obtained from multiple plane wave acquisitions steered under different angles \cite{MontaldoG2009Cpcf}, although this lowers the frame rate by an order of magnitude. Various methods have been considered to increase the image quality of plane wave imaging without decreasing the frame rate, such as adaptive minimum variance techniques, phase coherence imaging techniques, and sparse regularization methods \cite{KimKyuhong2014Afmv, ZhangJingke2021Uirf, CruzaJ.F2017Ppif}. While sparse regularization methods have been able to achieve high quality images, they have a large computational complexity in memory and time which makes them impractical.
Adaptive minimum variance techniques aim to suppress unwanted signals through apodization, but the methods also increase the computational complexity.
Phase coherence imaging techniques also aim to suppress unwanted signals by weighting the images with a phase coherence factor.
The method is however rather unstable due to coherence losses from shadowing and interference which will result in loss in quality.
Ultimately there remains a desire for improving the image quality of plane wave ultrasound imaging without decreasing the frame rate.

Deep learning is playing an ever growing role in ultrasound imaging as it has shown a lot of promise in tasks such as image processing, object detection, and image segmentation \cite{vanSlounRuudJ.G2020DLiU}.
Deep learning is also being applied in the physics-based image formation process for plane wave ultrasound imaging, in which it has been shown to be successful \cite{NairArunAsokan2018ADLB, StrohmHannah2020Dlro}.
Further investigation is however required to understand the additional potential that lies in the full information of the receive channel data.
To this end, various promising networks have been investigated, such as: convolutional neural networks for plane wave compounding \cite{GasseMaxime2017HPWC}, spatial coherence based beamforming using fully connected networks \cite{WiacekAlycen2020CADL}, and a fully convolutional network designed to model the united sign coherence factor \cite{YangChen2020AUSC, HyunDongwoon2021DLfU}.
A novel method that distinguishes itself by using physics-based deep learning has been proposed for single plane wave imaging in \cite{Pilikos_2021}. The core idea was to implement the $f$-$k$ migration algorithm for plane wave ultrasound imaging as a network layer to obtain high quality images from a single plane wave acquisition with an end-to-end data-to-image network. In \cite{Pilikos_2021}, the method was tested on a simple, simulated data scenario and showed the ability to locate inclusions accurately. In this work, we provide a thorough evaluation of the method with experimental data and compare it to more conventional approaches to utilize neural networks for plane wave ultrasound imaging. For the evaluation, we obtained an extensive experimental data collection that can be used for algorithm development and benchmarking in ultrasonic plane wave imaging. It consists of data acquired from a realistic breast mimicking phantom which can be used for network training and testing and data acquired from a multi-purpose calibration phantom. 
For the first part of the evaluation, we use the breast phantom data to measure global image similarity using image-to-image comparison metrics such as $\ell_{1}$ and $\ell_{2}$ loss, peak signal-to-noise ratio (PSNR), and normalized cross correlation (NCC).
In the second part of the evaluation, we use the calibration phantom to measure image quality relevant to ultrasound applications, such as contrast, resolution, and lesion detectability.

\section{Materials and Methods}

\begin{figure}[ht]
    \centering
    \subfloat[][]{\includegraphics[width=0.16\textwidth]{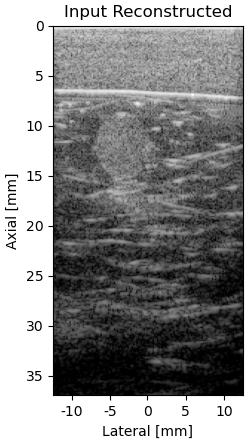}}
    \subfloat[][]{\includegraphics[width=0.16\textwidth]{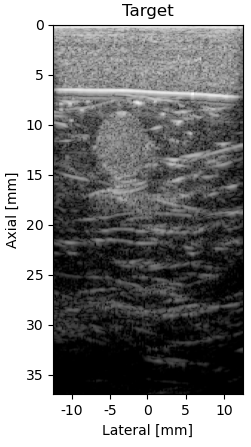}}
    \caption{Input and target image of a test sample: 
    (a) input data directly reconstructed by $f$-$k$ migration;
    (b) target image reconstructed by $f$-$k$ migration with 75 angles of compounding.}
    \label{fig:sample_img_n10_input_target}
\end{figure}

\begin{figure*}[t]
    \centering
    \includegraphics[width=1.0\textwidth]{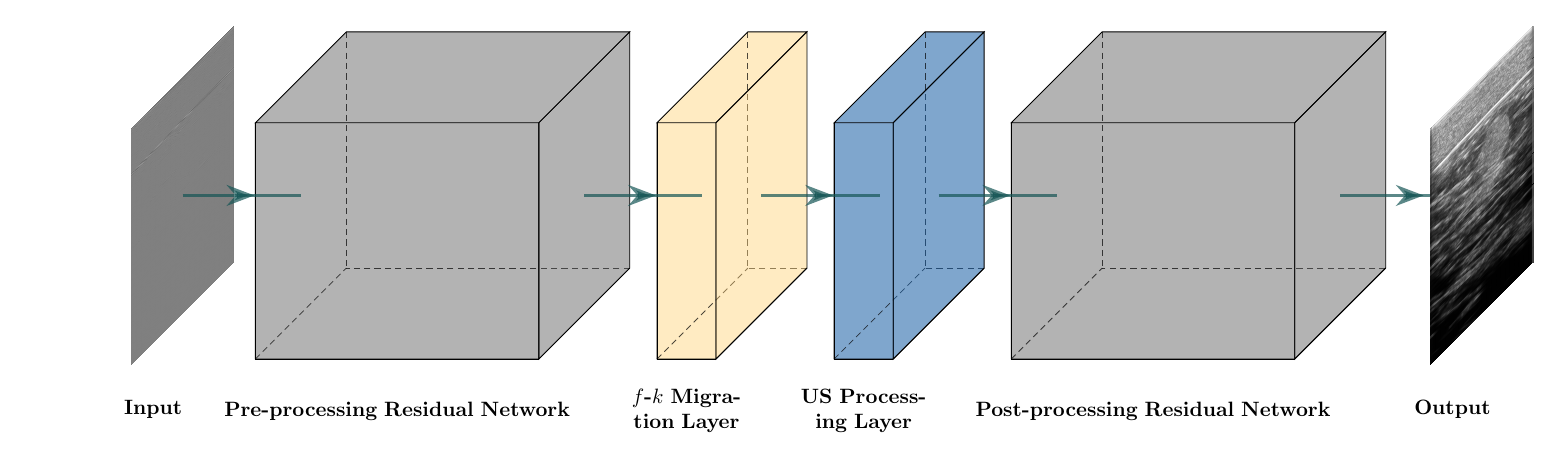}
    \caption{Overview of the proposed network architecture for the complete model.
    Leftmost is the raw channel data which serves as the input for the network.
    The first part of the network is a 2D residual network (ResNet) which acts as a pre-processing network on the raw channel data.
    The next step is the $f$-$k$ migration layer and the ultrasound image processing layer.
    The ultrasound processing layer consists of envelope detection, log compression and clipping.
    The final part of the network is another 2D ResNet, which acts as a post-processing layer on the image data.
    Finally the model output is obtained.}
    \label{fig:complete_model}
\end{figure*}

\subsection{Benchmarking Data Set Acquisition}

We acquired a data collection designed for training deep neural networks consisting of 220 ultrasound plane wave data recordings using an in-vitro breast-mimicking phantom. The breast phantom gives samples containing hyperechoic lesions, hypoechoic lesions, and normal breast tissue. Additional data was acquired from a calibration phantom ("Multi-Purpose Multi-Tissue Ultrasound Phantom Model 040GSE, CIRS, Norfolk, USA) to test resolution, contrast, and generalizability of trained networks: 5 samples of hypoechoic cysts, 5 samples from wire targets of 100$\mu$m, 5 samples with -6dB and -3dB lesions in the field of view, and 5 samples with +3dB and +6dB in the field of view. All recordings were obtained twice: once in a low attenuating area 0.7 dB/cm/mHz,  and once in a high attenuating area 0.95 dB/cm/mHz. 
All data acquisition was done using a Verasonics Vantage with 256 transmit and receive channels system equipped with a L12-5 50mm linear array ultrasound transducer operating at a center frequency of 7.8 MHz. No time-gain compensation was used. The plane wave acquisitions consisted of 75 steering angles with an angle range of $\left[-16, +16\right]$ degrees. 
We published the data set on Zenodo \cite{USdataset} alongside with codes for read-in and preprocessing (see Section \ref{subsec:CodeData}).

\subsection{The $f$-$k$ Migration for Plane Wave Image Formation}
The $f$-$k$ migration algorithm has its origin in seismic imaging, and it is based on the exploding reflector model (ERM) \cite{NumMeth}. Within the ERM model it is assumed that all scatters lie in the half-space $z > 0$ (axial direction) and ``explode'' at the same time $t = 0$ to become acoustic sources. The resulting scalar wave-field $\psi(x, z, t)$ is recorded at the surface $z=0$ over time. The $f$-$k$ migration algorithm then migrates the data $\psi(x, z = 0, t)$ back into the medium to reconstruct $\psi(x, z, t = 0)$. The migration is computed in Fourier domain, which leads to very efficient numerical implementations. In \cite{GarciaD2013Sfmf}, the ERM model and the $f$-$k$ migration algorithm were adapted to ultrasound imaging with tilted plane waves. Plane waves from multiple steering angles are averaged to form the final image (``angular compounding''). For this work, it is important to note that the numerical implementation of the $f$-$k$ migration algorithm we consider consists of very simple computational building blocks (cf. Fig. 5 in  \cite{GarciaD2013Sfmf} for more details): Fast Fourier transforms (\textit{fft}) with zero-padding along different dimensions, interpolation in Fourier domain and pointwise affine-linear transformations. 

After the $f$-$k$ migration step, ultrasound images are commonly further processed by an envelope detection using the Hilbert transform, log compression for better visualization, and clipping to get to the preferred dynamic range. Figure \ref{fig:sample_img_n10_input_target} (a) and (b) show images reconstructed with a single plane wave vs. plane waves from 75 angles steering angles from  breast-phantom data set. The 75 angle images will be the target images for training the neural networks described in the following section.

\subsection{Network Architectures}
In this work, all steps of the $f$-$k$ migration algorithm for plane wave ultrasound and the following post-processing steps were implemented in PyTorch as differentiable network layers using PyTorch's AutoGrad functionality \cite{NEURIPS2019_9015}.
Consequently, a neural network can be trained end-to-end with the $f$-$k$ migration and US-specific post-processing as part of its architecture.
The $f$-$k$ migration layer serves as an explicit mapping from the receive channel data to the image, so that the network does not need to learn this transformation.
In the same spirit, we chose to add the post-processing steps to the network explicitly instead of relying on part of the network's capacity and training data to learn them. 
The main network architecture consists of two 2D residual networks (ResNets) with the $f$-$k$ migration and image processing layers in between, see Fig. \ref{fig:complete_model}.
The first ResNet takes the raw channel data as input and outputs data of the same dimension. We therefore see it as a data-to-data pre-processing network. The second ResNet acts on the image reconstructed by the $f$-$k$ migration and US image processing layers. We therefore see it as an image-to-image post-processing network.
While we will refer to the resulting data-to-image network as the \textit{complete model} in the following, we will consider two variants of it: The \textit{post-processing model} only uses the image-to-image post-processing network to enhance the image reconstructed by the $f$-$k$ migration algorithm. Practically, this is implemented by omitting the pre-processing ResNet from the main architecture described in Fig. \ref{fig:complete_model}. For a fair comparison, the post-processing ResNet is twice as deep so that the model has the same number of trainable parameters and the same expressivity. The second variant, the \textit{pre-processing model}, in turn only uses a twice-as-deep pre-processing ResNet.

\begin{figure*}[t]
    \centering
    \includegraphics[width=1.0\textwidth]{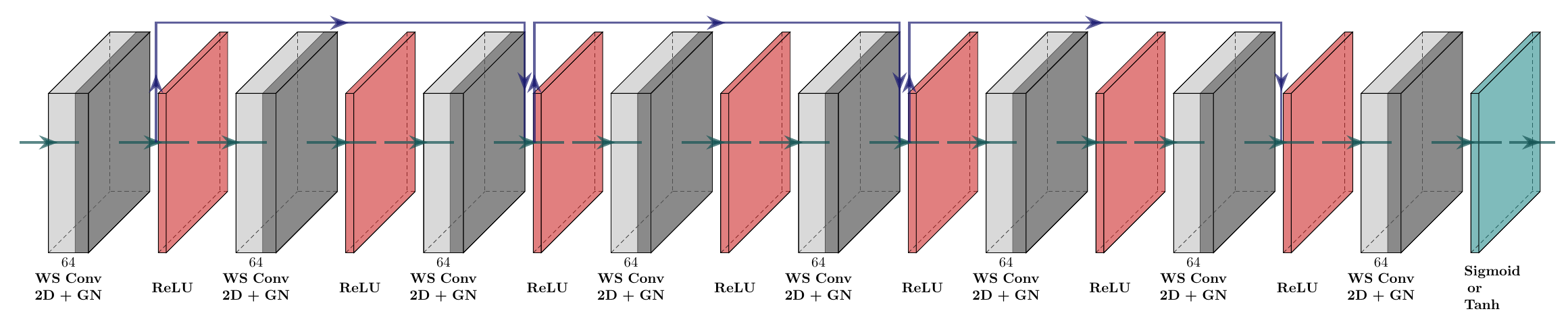}
    \caption{Graphical display of the ResNet used in the complete model depicted in Fig. \ref{fig:complete_model}.
    The ResNet is built up out of three residual blocks, which can be identified by the positioning of the skip connections.
    These residual blocks are comprised of 2D convolutional layers with weight standardization (WS) and group normalization (GN).
    These 2D convolutional layers use a kernel of size $5 \times 5$ with 64 channels per layer.
    Each layer except the last one uses a Rectified Linear Unit (ReLU) activation function.
    For the final activation function it holds that the pre-processing ResNet uses Tanh, and the post-processing ResNet uses Sigmoid.
    }
    \label{fig:ResNet_details}
\end{figure*}

The details of the ResNets used are shown in Fig. \ref{fig:ResNet_details}. They consist of three residual blocks, where the skip connections form the encompassment of these residual blocks.
The residual blocks consist of 2D convolutional layers with weight standardization (WS) and group normalization (GN) to increase training stability because a small batch size of 1 is used due to memory limitations \cite{QiaoSiyuan2019MTwB, WuYuxin2019GN}.
For the group normalization a number of 8 groups was used with learnable per-channel affine transform parameters.
Furthermore there are 16 convolutional layers, the convolutional layers have 64 channels per layer and use a convolutional kernel of size $5 \times 5$ with zero padding around the edges.

\subsection{Network Training}

The target images were constructed with the $f$-$k$ migration algorithm using 75 angles of compounding.
The raw channel data of single horizontal plane waves was used as the network input data. All neural networks were trained with the mean squared error loss function (see following section, Eqn \eqref{eq:l2loss}) for 70 epochs with a learning rate of 0.01 using the AMSGrad variant of the Adam optimizer with $\left(\beta_{1}, \beta_{2}\right) = \left(0.9, 0.999\right)$. From the 220 breast-mimicking samples, the same 44 samples were always used for testing, and the remaining 176 were available for training. The networks were trained with random sub-sets of these 176 samples, of sizes 4\%, 6\%, 8\%, and 10\%-100\% in steps of 10\%. In choosing the sub-sets, it was accounted for that each lesion variant was evenly represented in the training data for each percentage. Furthermore, for a fixed seed of the random number generator, smaller sub-sets are contained in larger sub-sets (e.g., all samples contained in the 4\% sub-set are also contained in the 6\% sub-set and so on). Each of the three network models was trained for each training data sub-set size using 16 different random seeds. Besides being trained on different training data samples, this randomization also leads to different weight initializations and mini-batch selection. 

\subsection{Image Quality Evaluation Metrics}
The evaluation metrics that were used can be categorized in three groups: Global image quality metrics serve as a direct image-to-image comparison of the output of the network architectures with the target images. Local image quality metrics aim to evaluate contrast and lesion detectability. Finally, the resolution is measured in terms of the full width at half maximum (FWHM).

\subsubsection{Global Image Quality Metrics}
Four global image-to-image quality metrics were considered: $\ell_{1}$ loss, $\ell_{2}$ loss, peak signal-to-noise ratio (PSNR), and the normalized cross correlation (NCC), defined as follows:
\begin{align}
    \ell_{1}\left(x, y\right) &= \frac{1}{N} \cdot \sum_{n = 1}^{N}\left|x_{n} - y_{n}\right|, \\
    \ell_{2}\left(x, y\right) &= \sqrt{\frac{1}{N} \cdot \sum_{n = 1}^{N}\left(x_{n} - y_{n}\right)^{2}}, \label{eq:l2loss} \\
    \text{PSNR}\left(x, y\right) &= 20 \cdot \text{log}_{10}\left(\frac{\text{max}_{n}\left(y_{n}\right)}{\ell_{2}\left(x, y\right)}\right), \\
    \text{NCC}\left(x, y\right) &= \frac{\displaystyle\sum_{n = 1}^{N} \left(x_{n} - \mu_{x} \right) \cdot \left(y_{n} - \mu_{y} \right)}{\sqrt{\left(\displaystyle\sum_{n = 1}^{N} \left(x_{n} - \mu_{x}\right)^2 \right) \cdot \left(\displaystyle\sum_{n = 1}^{N} \left(y_{n} - \mu_{y}\right)^2 \right)}}.
\end{align}
Here, $y$ represents the target image, and $x$ the image to evaluate, the subscript $n$ indexes the $n^{\text{th}}$ pixel and $N$ denotes the total number of pixels. The average intensity values are denoted by $\mu_{y}$ and $\mu_{x}$ respectively.
For $\ell_{1}$ loss and $\ell_{2}$ loss, lower values indicate better performance, whereas for PSNR and NCC, the opposite holds. Note that the $\ell_{2}$ loss is used as the loss function for network training.

\subsubsection{Local Image Quality Metrics}
Three local image quality metrics were considered: contrast ratio (CR), contrast-to-noise ratio (CNR), and the generalized contrast-to-noise ratio (gCNR) \cite{Rodriguez-MolaresAlfonso2020TGCR}.
These metrics compare statistics from two image regions in order to evaluate contrast and lesion detectability as:
\begin{align}
    \text{CR} &= 20 \cdot \text{log}_{10}\left(\frac{\mu_{1}}{\mu_{2}}\right), \\
    \text{CNR} &= 20 \cdot \text{log}_{10}\left(\frac{\left|\mu_{1} - \mu_{2} \right|}{\sqrt{\left(\sigma_{1}^{2} + \sigma_{2}^{2}\right)/ 2}}\right), \\
    \text{gCNR} &= 1 - \int_{-\infty}^{\infty} \text{min}\left\{h_1\left(s\right), h_2\left(s\right)\right\} \text{d}s, \nonumber \\
    &\approx 1 - \sum_{s} \text{min}\left\{h_1\left(s\right), h_2\left(s\right)\right\} \cdot \Delta s. \label{eq:gcnr2}
\end{align}
Here $\mu_{i}$ represents the mean signal, $\sigma_{i}^{2}$ the variance, and $h_{i}$ histograms as probability density functions in region $i = 1, 2$.
The gCNR metric was calculated with 256 histogram bins for the probability density functions $h_{i}$.

\subsubsection{Resolution}
The full width at half maximum (FWHM) was determined as a measure of resolution in both the axial and the lateral dimension from wire targets with a diameter of 100 $\mu$m. First, a ROI was selected around each wire target. Then, the average intensity profile within the ROIs for each dimension was calculated and a Gaussian was fitted to it. The FWHM was determined as $2 \sqrt{2 \cdot \log(2)} \sigma$, where $\sigma$ is the standard deviation of the Gaussian fit. 

\subsection{Code and Data Availability} \label{subsec:CodeData}

We provided Python and Matlab codes on github\footnote{\url{https://github.com/RALSchoop/DeepUS}} to access and pre-process the data collection \cite{USdataset}, compute reference reconstructions, set-up and train the network models using PyTorch \cite{NEURIPS2019_9015}, and evaluate all image quality metrics introduced above. As such, they provide a framework that researchers can use to benchmark plane wave image reconstruction methods. The $f$-$k$ migration implementation is based on \footnote{\url{https://github.com/rehmanali1994/Plane_Wave_Ultrasound_Stolt_F-K_Migration.github.io}}.

\section{Results}

\subsection{Breast Phantom Test Set Evaluation}

\begin{figure}[ht]
    \centering
    \subfloat[][]{\includegraphics[width=0.16\textwidth]{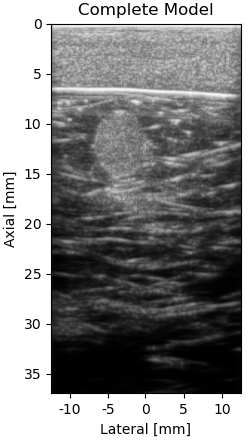}}
    \subfloat[][]{\includegraphics[width=0.16\textwidth]{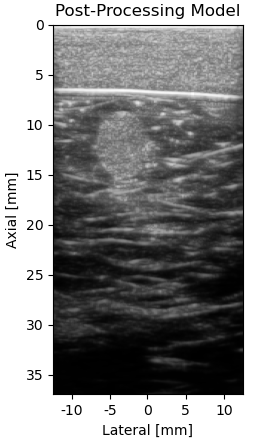}}
    \subfloat[][]{\includegraphics[width=0.16\textwidth]{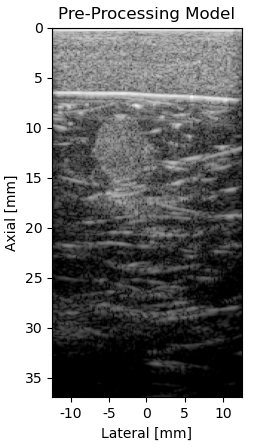}}
    \caption{Results on the same test sample as figure \ref{fig:sample_img_n10_input_target}: 
    (a) output complete model;
    (b) output post-processing model;
    (c) output pre-processing model.
    All networks were trained with 176 training samples.}
    \label{fig:sample_img_n10}
\end{figure}

Figure \ref{fig:sample_img_n10} shows a direct reconstruction of the single plane wave data from one test sample of the breast phantom, the corresponding reconstruction from 75 plane waves (which is the target in network training), and the output of all three networks models. The complete and post-processing model outputs look smoother overall when compared to the reconstructed input and target, contrary to the pre-processing model output which retains much more of this granular speckle in the image. Qualitatively, Figure \ref{fig:sample_img_n10} shows improved contrast for all model outputs when compared to the reconstructed input, although to a different degree.

\begin{figure}[ht]
    \centering
    \subfloat[][]{\includegraphics[width=0.24\textwidth]{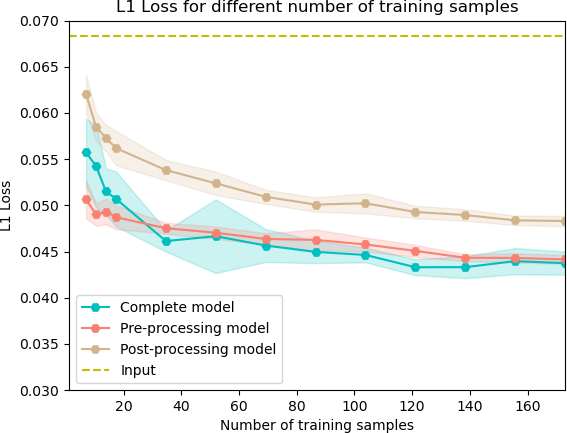}}
    \hspace{0.000001\textwidth}
    \subfloat[][]{\includegraphics[width=0.24\textwidth]{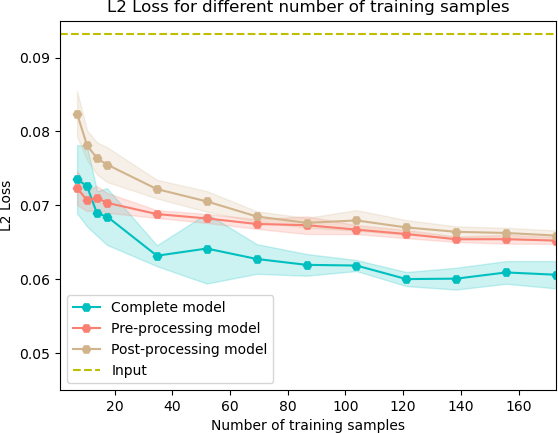}} \\
    \subfloat[][]{\includegraphics[width=0.24\textwidth]{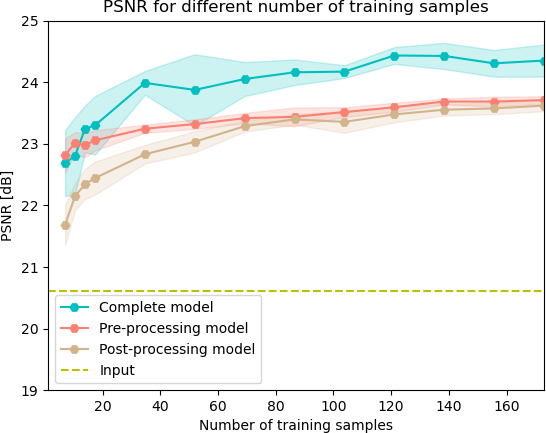}}
    \hspace{0.000001\textwidth}
    \subfloat[][]{\includegraphics[width=0.24\textwidth]{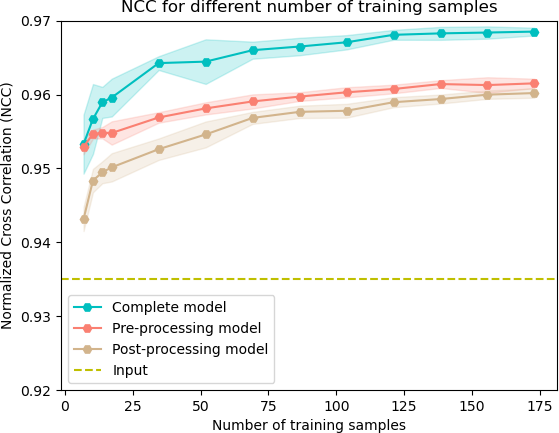}}
    \caption{Global image metrics calculated for the sample shown in Figure \ref{fig:sample_img_n10} as a function of the amount of training samples used. Shown are mean values $\pm$ standard deviation for the 16 random repetitions of the training:  (a) $\ell_{1}$ loss, (b) $\ell_{2}$ loss, (c) peak signal-to-noise ratio (PSNR), (d) normalized cross correlation (NCC).}
    \label{fig:global_metrics_n10}
\end{figure}

Figure \ref{fig:global_metrics_n10} shows the global evaluation metrics: $\ell_{1}$ loss, $\ell_{2}$ loss, peak signal-to-noise ratio (PSNR), and normalized cross correlation (NCC) as a function of the number of training samples calculated for the sample shown in Figure \ref{fig:sample_img_n10}. 
The results have been averaged over the 16 random repetitions (see above), and the plots display mean and standard deviation.  
Note that the target is used as the ground truth for the calculation of these global evaluation metrics, which is why it is not shown as a reference in Figure \ref{fig:global_metrics_n10}. 
The first observation to make is that all model outputs perform better than the input on all global evaluation metrics.
More interestingly this holds for all number of training samples, which implies improved performance with very little training data, e.g. 4\% amounts to only 7 training samples.
The performance improvement when adding more training data becomes quite small from 60 training samples onward.
From the three network variants, the complete model yields the best performance overall, surpassing the other models from about 20 training samples onward in all global evaluation metrics. Finally, one observes the post-processing model as performing the worst over all global evaluation metrics, albeit not with a large difference.

\subsection{Calibration Phantom Evaluation}

\begin{figure}[ht]
    \centering
    \subfloat[][]{\includegraphics[width=0.16\textwidth]{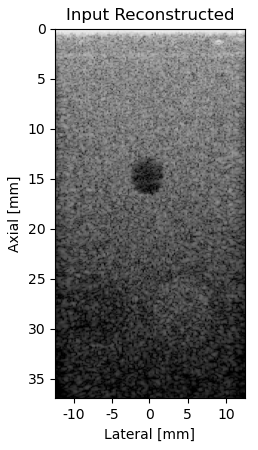}}
    \subfloat[][]{\includegraphics[width=0.16\textwidth]{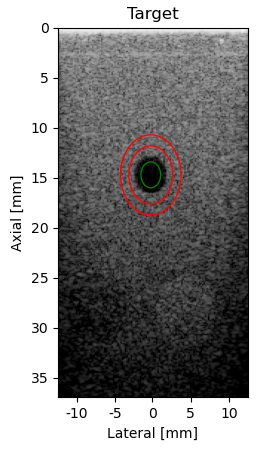}} \\
    \subfloat[][]{\includegraphics[width=0.16\textwidth]{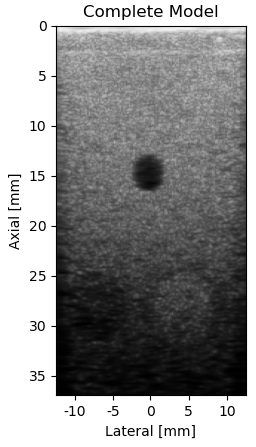}}
    \subfloat[][]{\includegraphics[width=0.16\textwidth]{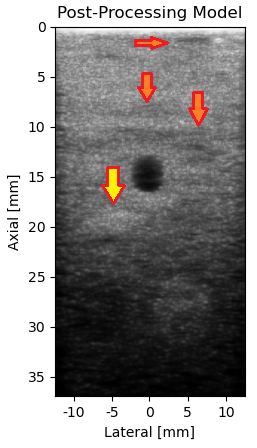}}
    \subfloat[][]{\includegraphics[width=0.16\textwidth]{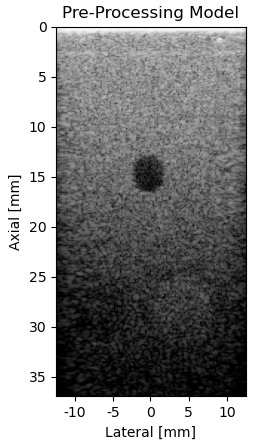}}
    \caption{Results on a hypo echoic lesion in the calibration phantom:
    (a) input data directly reconstructed by $f$-$k$ migration;
    (b) target image reconstructed by $f$-$k$ migration with 75 angles of compounding, including depiction of the region of interests (ROIs) used for the calculation of the local evaluation metrics;
    (c) output complete model;
    (d) output post-processing model, with arrows indicating regions with false intensity variations:  the yellow arrow indicates a false hyper echoic region while the orange arrows indicate false hypo echoic regions;
    (e) output pre-processing model; All networks were trained with 176 training samples.}
    \label{fig:img_n22}
\end{figure}

Contrast ratio (CR), contrast-to-noise ratio (CNR), and generalized contrast-to-noise ratio (gCNR) were calculated from the hypo echoic lesion in the calibration phantom in order to assess image contrast quantitatively, cf. Figure  \ref{fig:img_n22} (a, b). However, the results also contain an important qualitative observation about the post-processing model: Figure \ref{fig:img_n22}(d) shows that the post-processing model hallucinates false inclusions into the image. The yellow arrow shows a clear instance of a false hyper echoic region and the orange arrows show some clear instances of false hypo echoic regions. 
These inclusions are not present in the reconstructed input or the target and are therefore an unwanted artifact of this model variant. The post-processing model is the only network variant which introduces these artifacts.

\begin{figure}[ht]
    \centering
    \subfloat[][]{\includegraphics[width=0.25\textwidth]{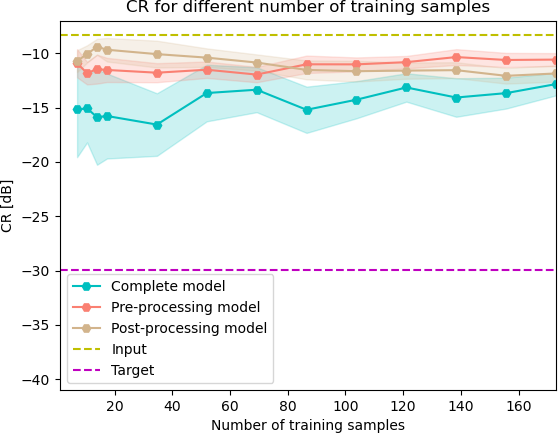}} \\
    \subfloat[][]{\includegraphics[width=0.24\textwidth]{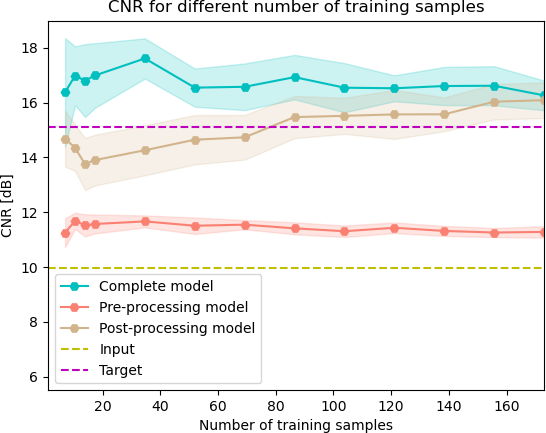}}
    \hspace{0.000001\textwidth}
    \subfloat[][]{\includegraphics[width=0.24\textwidth]{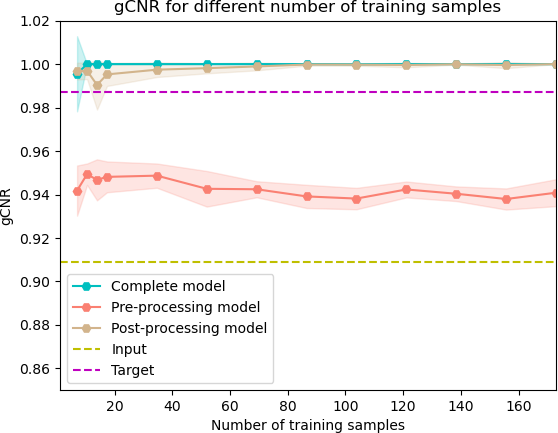}}
    \caption{Local image metrics as a function of the amount of training samples used, computed for the ROIs depicted in Figure \ref{fig:img_n22}(b). Shown are mean values $\pm$ standard deviation for the 16 random repetitions of the training: (a) contrast ratio (CR), (b) contrast-to-noise ratio (CNR), (c) generalized contrast-to-noise ratio (gCNR).}
    \label{fig:local_metrics_n22}
\end{figure}

Figure \ref{fig:local_metrics_n22} shows the CR, CNR, and gCNR as function of the number of training samples. 
Since the lesion is hypo echoic, the CR is negative and it holds that the lower the value the better the CR.
The target shows a contrast ratio of -30 dB. While the model outputs do not reach this value, they all improve over the input with the complete model and the post-processing model performing the best with a CR of around -14 dB at the maximum number of training samples.
In terms of the CNR and gCNR it is striking that the pre-processing model is not able to match the same level of performance as the complete and post-processing model outputs.
On the other hand, the post-processing model and complete model are outperforming the target in terms of CNR and gCNR for a sufficient number of training samples. 
In case of the complete model the target is outperformed for every number of training samples.
Interestingly, a gCNR of 1 is attained for these model outputs, which means a complete separation of the intensity histograms in the different regions.

\subsection{Resolution}
\begin{figure}[ht]
    \centering
    \subfloat[][]{\includegraphics[width=0.16\textwidth]{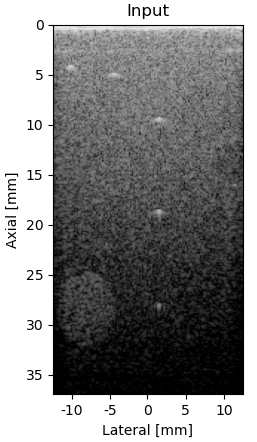}}
    \subfloat[][]{\includegraphics[width=0.158\textwidth]{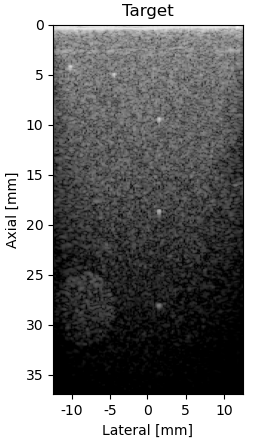}} \\
    \subfloat[][]{\includegraphics[width=0.16\textwidth]{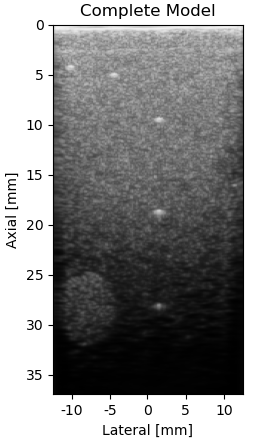}}
    \subfloat[][]{\includegraphics[width=0.16\textwidth]{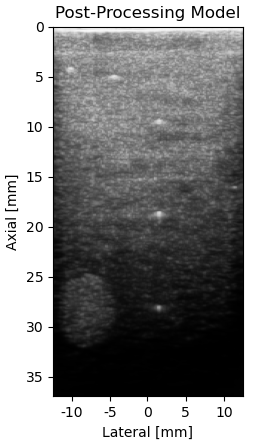}}
    \subfloat[][]{\includegraphics[width=0.16\textwidth]{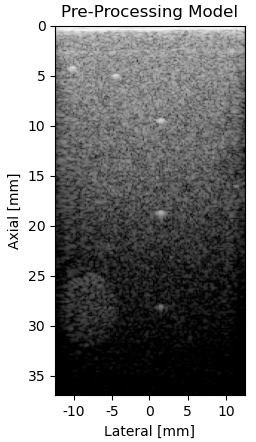}}
    \caption{Results on 100 $\mu\text{m}$ wire targets in the calibration phantom:
    (a) input data directly reconstructed by $f$-$k$ migration;
    (b) target image reconstructed by $f$-$k$ migration with 75 angles of compounding;
    (c) output complete model;
    (d) output post-processing model, similar to figure \ref{fig:img_n22}(d), there are false intensity variations in the generated image;
    (e) output pre-processing model; All networks were trained with 176 training samples.}
    \label{fig:img_n6}
\end{figure}

\begin{figure}[ht]
    \centering
    \subfloat[][]{\includegraphics[width=0.24\textwidth]{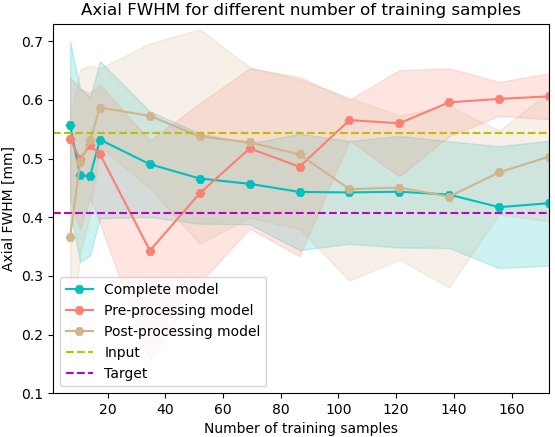}}
    \hspace{0.000001\textwidth}
    \subfloat[][]{\includegraphics[width=0.24\textwidth]{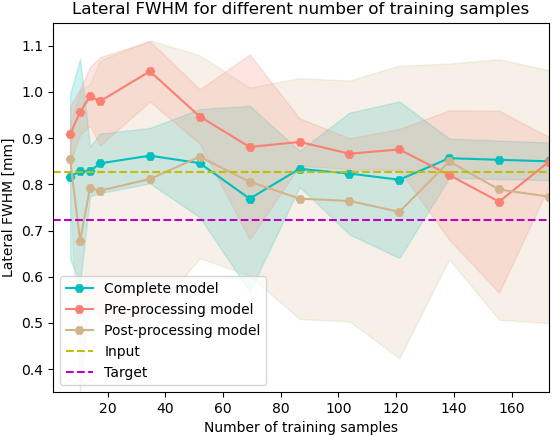}}
    \caption{Full width at half maximum (FWHM) as a function of the number of training samples. Shown are mean values $\pm$ standard deviation for the 16 random repetitions of the training for (a) the axial dimension and (b) the lateral dimension.}
    \label{fig:res_n6}
\end{figure}

In Figure \ref{fig:img_n6} shows all reconstructed images for the wire section of the calibration phantom.
Notice again how the post-processing model is hallucinating false hyper echoic and hypo echoic regions in the reconstructed image.
In Figure \ref{fig:res_n6} the image resolution is measured in terms of the full width at half maximum (FWHM) as a function of the number of training samples for all the model variants. While the complete model seems to improve upon the axial resolution with growing number of training samples, no model seems to achieve  clear improvements in lateral resolution.

\section{Discussion}
In this work, a deep neural network architecture that maps single plane wave US data to an image by using an explicit image formation layer was evaluated in-depth. For this, experimental data sets from a realistic breast phantom were collected in a way that allows to train and benchmark deep-learning-based plane wave US imaging methods.  Overall, the proposed network outperformed network variants that only work in either data or image domain regarding the global and local evaluation metrics considered, cf. Figure \ref{fig:global_metrics_n10} and \ref{fig:local_metrics_n22}. In terms of resolution however, the network could only provide moderate improvements in axial resolution compared to the reconstructed input, see Figure \ref{fig:res_n6}.

Most notably, all evaluation metrics already show improved performance of the model outputs compared to the reconstructed input with as little as 4\%, i.e., 7 samples, of the total amount of available training data. 
These results imply that the models considered here are able to learn quite robustly with little amounts of data. 
An important point of future work is to extend this evaluation to data-to-image networks without explicit image formation layer, i.e., networks that learns the transformation from data to image domain completely data-driven, e.g. as presented in \cite{segmNair,DahanElay2023DMUB}.

Figures \ref{fig:img_n22}(d) and \ref{fig:img_n6} (d) show a striking difference between the models considered: When evaluating the models trained on the breast phantom on the calibration phantom, the image-to-image post-processing network hallucinates arbitrary hyper and hypo intensity regions into the images which look like lesions. This could be an indication that the network is overfitting to the data set used for training and testing. The training samples of the breast phantom all contain some hyper-echoic and hypo-echoic regions and contains many small scale image features throughout the image. In contrast, the calibration phantom has a more homogeneous background except for where the specific inclusions are. While the image artifacts highlighted in Figures \ref{fig:img_n22}(d) and \ref{fig:img_n6} (d) are most undesirable in US applications, they are not sufficiently represented in the global evaluation metrics that we currently used. A way to improve upon this is to use window based evaluation metrics, resulting in a similarity map. To this end the multi-scale structural similarity index measure (MS-SSIM), or the window based variant of the normalized cross correlation could be used.

From Figure \ref{fig:sample_img_n10}, it is observed that the complete model and post-processing model are producing images that are smoother than the reconstructed input and target, whereas the pre-processing method does not show this. In other words, the output image is smoothed for the models that have an image-to-image 2D ResNet in their architecture, which implies that the smoothing happens inside the image-to-image ResNet.
The smoothing ultimately has a denoising effect, and Figure \ref{fig:local_metrics_n22} shows that the contrast metrics are improved. However, this effect might actually be undesirable in certain ultrasound applications, because it changes the noise-like speckle pattern which can contain critical information.
A next step of research would be to investigate the properties of the speckle pattern of the proposed models.

Figure \ref{fig:img_n22}(b) depicts the ROIs taken for the calculation of the local image quality metrics.
One problem of choosing such an ROI pair is that they are not at the same depth: The ring has a part that is above the circle and a part that is below the circle. This may systematically affect the calculated mean and variance for the evaluation metrics, because of the intensity loss of ultrasound with increasing depth. A way to account for this would be to fit an intensity decaying model as function of depth and correct the loss in intensity using it. Another possibility is to use time-gain compensation as is commonly done in ultrasound applications. The use of this time gain compensation, however enhances the noise in regions that are deeper in the medium, which can still systematically affect the calculated statistics for the evaluation.

Figures \ref{fig:img_n6} and \ref{fig:res_n6} show that the considered models can only provide moderate improvements with respect to the axial resolution of a point scatterer. This was however, also not to be expected: The training data set did not contain any point source samples and the $\ell_2$ training loss \eqref{eq:l2loss} is well know to lead to networks that rather smooth than sharpen the images, which is desirable for obtaining a good contrast (cf. Figure \ref{fig:local_metrics_n22}). Research into this phenomena and the development of alternative loss functions is a an active field of research (see Section 3.5.1. in \cite{LEPCHA2023230} for a recent overview and \cite{ZhGaFrKa17,BlMi18}). An important direction of future work for US imaging is to develop loss functions that are most sensitive to those image features that are most relevant for the clinical task at hand.

\section{Conclusion}

The proposed physics-based deep learning models improve the quality of single plane wave imaging on all global metrics and in terms of contrast with small amounts of experimental training data. It outperforms the more conventional image-to-image post-processing network and the data-to-image pre-processing network. The method however is currently unable to increase the resolution of single plane wave imaging.

\ifCLASSOPTIONcaptionsoff
  \newpage
\fi

\bibliographystyle{IEEEtran}
\bibliography{references}

\begin{thebibliography}{10}
\providecommand{\url}[1]{#1}
\csname url@samestyle\endcsname
\providecommand{\newblock}{\relax}
\providecommand{\bibinfo}[2]{#2}
\providecommand{\BIBentrySTDinterwordspacing}{\spaceskip=0pt\relax}
\providecommand{\BIBentryALTinterwordstretchfactor}{4}
\providecommand{\BIBentryALTinterwordspacing}{\spaceskip=\fontdimen2\font plus
\BIBentryALTinterwordstretchfactor\fontdimen3\font minus
  \fontdimen4\font\relax}
\providecommand{\BIBforeignlanguage}[2]{{%
\expandafter\ifx\csname l@#1\endcsname\relax
\typeout{** WARNING: IEEEtran.bst: No hyphenation pattern has been}%
\typeout{** loaded for the language `#1'. Using the pattern for}%
\typeout{** the default language instead.}%
\else
\language=\csname l@#1\endcsname
\fi
#2}}
\providecommand{\BIBdecl}{\relax}
\BIBdecl

\bibitem{SzaboThomasL2014Duii}
T.~L. Szabo, \emph{\BIBforeignlanguage{eng}{Diagnostic ultrasound imaging:
  inside out, second edition}}, 2nd~ed.\hskip 1em plus 0.5em minus 0.4em\relax
  Waltham, MA: Academic Press, 2014.

\bibitem{LuijtenBen2020AUBU}
B.~Luijten, R.~Cohen, F.~J. de~Bruijn, H.~A.~W. Schmeitz, M.~Mischi, Y.~C.
  Eldar, and R.~J.~G. van Sloun, ``\BIBforeignlanguage{eng}{Adaptive ultrasound
  beamforming using deep learning},'' \emph{\BIBforeignlanguage{eng}{IEEE
  transactions on medical imaging}}, vol.~39, no.~12, pp. 3967--3978, 2020.

\bibitem{AlbertiGiovanniS.2017MAoU}
G.~S. Alberti, H.~Ammari, F.~Romero, and T.~Wintz,
  ``\BIBforeignlanguage{eng}{Mathematical analysis of ultrafast ultrasound
  imaging},'' \emph{\BIBforeignlanguage{eng}{SIAM journal on applied
  mathematics}}, vol.~77, no.~1, pp. 1--25, 2017.

\bibitem{TanterMickael2014Uiib}
M.~Tanter and M.~Fink, ``\BIBforeignlanguage{eng}{Ultrafast imaging in
  biomedical ultrasound},'' \emph{\BIBforeignlanguage{eng}{IEEE transactions on
  ultrasonics, ferroelectrics, and frequency control}}, vol.~61, no.~1, pp.
  102--119, 2014.

\bibitem{BercoffJ2004Ssia}
J.~Bercoff, M.~Tanter, and M.~Fink, ``\BIBforeignlanguage{eng}{Supersonic shear
  imaging: a new technique for soft tissue elasticity mapping},''
  \emph{\BIBforeignlanguage{eng}{IEEE transactions on ultrasonics,
  ferroelectrics, and frequency control}}, vol.~51, no.~4, pp. 396--409, 2004.

\bibitem{ErricoClaudia2015Uulm}
C.~Errico, J.~Pierre, S.~Pezet, Y.~Desailly, Z.~Lenkei, O.~Couture, and
  M.~Tanter, ``\BIBforeignlanguage{eng}{Ultrafast ultrasound localization
  microscopy for deep super-resolution vascular imaging},''
  \emph{\BIBforeignlanguage{eng}{Nature (London)}}, vol. 527, no. 7579, pp.
  499--502, 2015.

\bibitem{MontaldoG2009Cpcf}
G.~Montaldo, M.~Tanter, J.~Bercoff, N.~Benech, and M.~Fink,
  ``\BIBforeignlanguage{eng}{Coherent plane-wave compounding for very high
  frame rate ultrasonography and transient elastography},''
  \emph{\BIBforeignlanguage{eng}{IEEE transactions on ultrasonics,
  ferroelectrics, and frequency control}}, vol.~56, no.~3, pp. 489--506, 2009.

\bibitem{KimKyuhong2014Afmv}
K.~Kim, S.~Park, J.~Kim, S.-B. Park, and M.~Bae, ``\BIBforeignlanguage{eng}{A
  fast minimum variance beamforming method using principal component
  analysis},'' \emph{\BIBforeignlanguage{eng}{IEEE transactions on ultrasonics,
  ferroelectrics, and frequency control}}, vol.~61, no.~6, pp. 930--945, 2014.

\bibitem{ZhangJingke2021Uirf}
J.~Zhang, Q.~He, Y.~Xiao, H.~Zheng, C.~Wang, and J.~Luo,
  ``\BIBforeignlanguage{eng}{Ultrasound image reconstruction from plane wave
  radio-frequency data by self-supervised deep neural network},''
  \emph{\BIBforeignlanguage{eng}{Medical image analysis}}, vol.~70, pp.
  102\,018--102\,018, 2021.

\bibitem{CruzaJ.F2017Ppif}
J.~Cruza, J.~Camacho, and C.~Fritsch, ``\BIBforeignlanguage{eng}{Plane-wave
  phase-coherence imaging for nde},'' \emph{\BIBforeignlanguage{eng}{NDT \& E
  international : independent nondestructive testing and evaluation}}, vol.~87,
  pp. 31--37, 2017.

\bibitem{vanSlounRuudJ.G2020DLiU}
R.~J.~G. van Sloun, R.~Cohen, and Y.~C. Eldar, ``\BIBforeignlanguage{eng}{Deep
  learning in ultrasound imaging},'' \emph{\BIBforeignlanguage{eng}{Proceedings
  of the IEEE}}, vol. 108, no.~1, pp. 11--29, 2020.

\bibitem{NairArunAsokan2018ADLB}
A.~A. Nair, T.~D. Tran, A.~Reiter, and M.~A. Lediju~Bell,
  ``\BIBforeignlanguage{eng}{A deep learning based alternative to beamforming
  ultrasound images},'' in \emph{\BIBforeignlanguage{eng}{2018 IEEE
  International Conference on Acoustics, Speech and Signal Processing
  (ICASSP)}}.\hskip 1em plus 0.5em minus 0.4em\relax IEEE, 2018, pp.
  3359--3363.

\bibitem{StrohmHannah2020Dlro}
H.~Strohm, S.~Rothlübbers, K.~Eickel, and M.~Günther,
  ``\BIBforeignlanguage{eng}{Deep learning-based reconstruction of ultrasound
  images from raw channel data},'' \emph{\BIBforeignlanguage{eng}{International
  journal for computer assisted radiology and surgery}}, vol.~15, no.~9, pp.
  1487--1490, 2020.

\bibitem{GasseMaxime2017HPWC}
M.~Gasse, F.~Millioz, E.~Roux, D.~Garcia, H.~Liebgott, and D.~Friboulet,
  ``\BIBforeignlanguage{eng}{High-quality plane wave compounding using
  convolutional neural networks},'' \emph{\BIBforeignlanguage{eng}{IEEE
  transactions on ultrasonics, ferroelectrics, and frequency control}},
  vol.~64, no.~10, pp. 1637--1639, 2017.

\bibitem{WiacekAlycen2020CADL}
A.~Wiacek, E.~Gonzalez, and M.~A.~L. Bell,
  ``\BIBforeignlanguage{eng}{Coherenet: A deep learning architecture for
  ultrasound spatial correlation estimation and coherence-based beamforming},''
  \emph{\BIBforeignlanguage{eng}{IEEE transactions on ultrasonics,
  ferroelectrics, and frequency control}}, vol.~67, no.~12, pp. 2574--2583,
  2020.

\bibitem{YangChen2020AUSC}
C.~Yang, Y.~Jiao, T.~Jiang, Y.~Xu, and Y.~Cui, ``\BIBforeignlanguage{eng}{A
  united sign coherence factor beamformer for coherent plane-wave compounding
  with improved contrast},'' \emph{\BIBforeignlanguage{eng}{Applied sciences}},
  vol.~10, no.~7, pp. 2250--, 2020.

\bibitem{HyunDongwoon2021DLfU}
D.~Hyun, A.~Wiacek, S.~Goudarzi, S.~Rothlubbers, A.~Asif, K.~Eickel, Y.~C.
  Eldar, J.~Huang, M.~Mischi, H.~Rivaz, D.~Sinden, R.~J.~G. van Sloun,
  H.~Strohm, and M.~A.~L. Bell, ``\BIBforeignlanguage{eng}{Deep learning for
  ultrasound image formation: Cubdl evaluation framework and open datasets},''
  \emph{\BIBforeignlanguage{eng}{IEEE transactions on ultrasonics,
  ferroelectrics, and frequency control}}, vol.~68, no.~12, pp. 3466--3483,
  2021.

\bibitem{Pilikos_2021}
\BIBentryALTinterwordspacing
G.~Pilikos, C.~L. de~Korte, T.~van Leeuwen, and F.~Lucka, ``Single plane-wave
  imaging using physics-based deep learning,'' in \emph{2021 {IEEE}
  International Ultrasonics Symposium ({IUS})}.\hskip 1em plus 0.5em minus
  0.4em\relax {IEEE}, sep 2021. [Online]. Available:
  \url{https://doi.org/10.1109%2Fius52206.2021.9593589}
\BIBentrySTDinterwordspacing

\bibitem{USdataset}
\BIBentryALTinterwordspacing
R.~A. Schoop, ``{Ultrasound Plane Wave Raw Data 75 Angles - Breast Phantom and
  Calibration Phantom Dataset},'' Jun. 2023. [Online]. Available:
  \url{https://doi.org/10.5281/zenodo.7986407}
\BIBentrySTDinterwordspacing

\bibitem{NumMeth}
\BIBentryALTinterwordspacing
G.~F. Margrave. (2003) Numerical methods of exploration seismology. [Online].
  Available:
  \url{https://www.crewes.org/ResearchLinks/FreeSoftware/NumMeth.pdf}
\BIBentrySTDinterwordspacing

\bibitem{GarciaD2013Sfmf}
D.~Garcia, L.~L. Tarnec, S.~Muth, E.~Montagnon, J.~Por\'{e}e, and G.~Cloutier,
  ``\BIBforeignlanguage{eng}{Stolt's f-k migration for plane wave ultrasound
  imaging},'' \emph{\BIBforeignlanguage{eng}{IEEE transactions on ultrasonics,
  ferroelectrics, and frequency control}}, vol.~60, no.~9, pp. 1853--1867,
  2013.

\bibitem{NEURIPS2019_9015}
\BIBentryALTinterwordspacing
A.~Paszke, S.~Gross, F.~Massa, A.~Lerer, J.~Bradbury, G.~Chanan, T.~Killeen,
  Z.~Lin, N.~Gimelshein, L.~Antiga, A.~Desmaison, A.~Kopf, E.~Yang, Z.~DeVito,
  M.~Raison, A.~Tejani, S.~Chilamkurthy, B.~Steiner, L.~Fang, J.~Bai, and
  S.~Chintala, ``Pytorch: An imperative style, high-performance deep learning
  library,'' in \emph{Advances in Neural Information Processing Systems
  32}.\hskip 1em plus 0.5em minus 0.4em\relax Curran Associates, Inc., 2019,
  pp. 8024--8035. [Online]. Available:
  \url{http://papers.neurips.cc/paper/9015-pytorch-an-imperative-style-high-performance-deep-learning-library.pdf}
\BIBentrySTDinterwordspacing

\bibitem{QiaoSiyuan2019MTwB}
S.~Qiao, H.~Wang, C.~Liu, W.~Shen, and A.~Yuille,
  ``\BIBforeignlanguage{eng}{Micro-batch training with batch-channel
  normalization and weight standardization},'' 2019.

\bibitem{WuYuxin2019GN}
Y.~Wu and K.~He, ``\BIBforeignlanguage{eng}{Group normalization},''
  \emph{\BIBforeignlanguage{eng}{International journal of computer vision}},
  vol. 128, no.~3, pp. 742--755, 2019.

\bibitem{Rodriguez-MolaresAlfonso2020TGCR}
A.~Rodriguez-Molares, O.~M.~H. Rindal, J.~D'hooge, S.-E. Masoy, A.~Austeng,
  M.~A. Lediju~Bell, and H.~Torp, ``\BIBforeignlanguage{eng ; nor}{The
  generalized contrast-to-noise ratio: A formal definition for lesion
  detectability},'' \emph{\BIBforeignlanguage{eng ; nor}{IEEE transactions on
  ultrasonics, ferroelectrics, and frequency control}}, vol.~67, no.~4, pp.
  745--759, 2020.

\bibitem{segmNair}
A.~A. {Nair}, K.~N. {Washington}, T.~D. {Tran}, A.~{Reiter}, and M.~A.~L.
  {Bell}, ``Deep learning to obtain simultaneous image and segmentation outputs
  from a single input of raw ultrasound channel data,'' \emph{IEEE Transactions
  on Ultrasonics, Ferroelectrics, and Frequency Control}, 2020.

\bibitem{DahanElay2023DMUB}
E.~Dahan and I.~Cohen, ``\BIBforeignlanguage{eng}{Deep-learning-based multitask
  ultrasound beamforming},'' \emph{\BIBforeignlanguage{eng}{Information
  (Basel)}}, vol.~14, no.~10, pp. 582--, 2023.

\bibitem{LEPCHA2023230}
\BIBentryALTinterwordspacing
D.~C. Lepcha, B.~Goyal, A.~Dogra, and V.~Goyal, ``Image super-resolution: A
  comprehensive review, recent trends, challenges and applications,''
  \emph{Information Fusion}, vol.~91, pp. 230--260, 2023. [Online]. Available:
  \url{https://www.sciencedirect.com/science/article/pii/S1566253522001762}
\BIBentrySTDinterwordspacing

\bibitem{ZhGaFrKa17}
H.~Zhao, O.~Gallo, I.~Frosio, and J.~Kautz, ``Loss functions for image
  restoration with neural networks,'' \emph{IEEE Transactions on Computational
  Imaging}, vol.~3, no.~1, pp. 47--57, 2017.

\bibitem{BlMi18}
Y.~Blau and T.~Michaeli, ``The perception-distortion tradeoff,'' in \emph{2018
  IEEE/CVF Conference on Computer Vision and Pattern Recognition}, 2018, pp.
  6228--6237.

\end{thebibliography}

\end{document}